\documentstyle[aps,epsfig,float,amstex]{revtex}
\newcommand{\be}{\begin{equation}}
\newcommand{\ee}{\end{equation}}
\newcommand{\beq}{\begin{eqnarray}}
\newcommand{\eeq}{\end{eqnarray}}
 
\begin{document}
    
\def\gC{\mbox{\boldmath $C$}}
\def\gZ{\mbox{\boldmath $Z$}}
\def\gR{\mbox{\boldmath $R$}}
\def\gN{\mbox{\boldmath $N$}}
\def\ua{\uparrow}
\def\da{\downarrow}
\def\a{\alpha}
\def\b{\beta}
\def\g{\gamma}
\def\G{\Gamma}
\def\d{\delta}
\def\D{\Delta}
\def\e{\epsilon}
\def\ve{\varepsilon}
\def\z{\zeta}
\def\h{\eta}
\def\th{\theta}
\def\k{\kappa}
\def\l{\lambda}
\def\L{\Lambda}
\def\m{\mu}
\def\n{\nu}
\def\x{\xi}
\def\X{\Xi}
\def\p{\pi}
\def\P{\Pi}
\def\r{\rho}
\def\s{\sigma}
\def\S{\Sigma}
\def\t{\tau}
\def\f{\phi}
\def\vf{\varphi}
\def\F{\Phi}
\def\c{\chi}
\def\w{\omega}
\def\W{\Omega}
\def\Q{\Psi}
\def\q{\psi}
\def\de{\partial}
\def\inf{\infty}
\def\ra{\rightarrow}
\def\bra{\langle}
\def\ket{\rangle}
\title{$W=0$ Pairing in $(N,N)$ Carbon Nanotubes away from 
Half Filling}
\author{Enrico Perfetto, Gianluca Stefanucci and Michele Cini}
\address{Istituto Nazionale per la Fisica della Materia, Dipartimento di Fisica,\\
Universita' di Roma Tor Vergata, Via della Ricerca Scientifica, 1-00133\\
Roma, Italy}
\maketitle

\begin{abstract}

We use the Hubbard Hamiltonian $H$ on the honeycomb lattice 
to represent the valence bands of carbon single-wall $(N,N)$ nanotubes. 
A detailed symmetry analysis shows that the model allows $W=0$ pairs 
which we define as two-body singlet eigenstates of $H$ with vanishing 
on-site repulsion. By means of a non-perturbative canonical transformation 
we calculate the effective interaction between the 
electrons of a $W=0$ pair added to the interacting ground state. 
We show that the dressed $W=0$ pair is a bound state for resonable 
parameter values away from half filling. Exact diagonalization results 
for the (1,1) nanotube confirm the 
expectations. For $(N,N)$ nanotubes of length $l$, the binding energy 
of the pair  depends strongly on the filling and decreases towards a small 
but nonzero value as $l \rightarrow \infty$. We observe the existence 
of an optimal doping when the number of electrons per C atom is in 
the range 1.2$\div$1.3, and the binding energy is of the  order of 
0.1 $\div$ 1 meV.

\end{abstract}
  
\bigskip
{\small 

\section{Introduction}

After the discovery of carbon nanotubes\cite{iijima} the interest 
in such systems has been 
stimulated by their anomalous normal properties\cite{bockrathetal} 
and by the recently reported superconductivity\cite{kasumov}. Indeed there has 
been growing evidence of superconducting fluctuations in single-wall carbon 
nanotubes placed between superconducting 
contacts\cite{morpurgo}\cite{tang}\cite{kociak} 
up to the transition temperature of $\simeq$ 0.5 K\cite{kasumov}.  

 A single-wall carbon nanotube (SWNT) is a graphite sheet wrapped onto 
a cylinder. The carbon atoms are arranged on the sites of a honeycomb 
lattice. The two primitive Bravais lattice vectors are ${\mathbf a}_{\pm }=
(d/2)(\pm 1,\sqrt{3})$, where $d/\sqrt{3}$ is the nearest neighbor carbon 
separation, see Fig.(\ref{honey}). A SWNT is characterized by a pair of 
integers $(N,M)$ which specifies the wrapping: the cylinder has the axis 
running perpendicular to $N{\mathbf a}_{+}+M{\mathbf a}_{-}$, so that atoms 
separated by $N{\mathbf a}_{+}+M{\mathbf a}_{-}$ are identified. Only 
in recent years, was it possible to study the electronic 
properties of atomically resolved SWNT's; it was found\cite{wildoer} that 
they are strongly dependent on the integers $N$ and $M$. 

From band-structure calculations\cite{hamada}\cite{blase} one predicts 
that   the ``armchair'' $(N,N)$ tubes are metals  while the ``zig-zag'' 
$(N,-N)$ ones [which are the same as the $(N,0)$ tubes] are insulators 
or semiconductors.  However, the Coulomb interaction cannot be 
neglected\cite{tans}\cite{bockrath}. 
Due to the quasi one-dimensional structure, the SWNT's are believed to 
exhibit non-Fermi liquid behaviour\cite{bellucci}. The band structure of 
the armchair nanotubes 
suggests\cite{balents}\cite{sol} that near half-filling the low-energy 
effective Hamiltonian should be a four-component Luttinger 
model. This theory was developed by the perturbative 
Renormalization Group\cite{balents}\cite{krotov} and by the 
bosonization technique\cite{gogolin}\cite{yo} in the weakly doped 
case, revealing the presence of a superconducting instability for 
short-ranged interactions. Still in a one-dimensional scheme, phonon 
induced pairing mechanisms 
have also been reported\cite{caron}\cite{gonzo}.

On the other hand, in this exploratory paper we consider a different 
scenario in which the electron density is so far from half-filling that 
any one-dimensional free-fermion model with a linear spectrum fails to  
describe the non-interacting Hamiltonian. This is motivated by the 
expectation\cite{kociak} that an increased doping by chemical manipulations 
and/or external fields could enhance the superconducting transition 
temperature. In recent years, we proposed a 
pairing-mechanism\cite{EPJB1999}\cite{EPJB2001} in the 
{\em two-dimensional} (one-band and three-bands) repulsive Hubbard 
model for the Cuprates.
Here we show that the same idea extends to the case at hand.  In our 
approach bound electron-pairs in SWNT are obtained by a symmetry-driven 
configuration interaction mechanism in which the transverse direction plays a 
crucial r$\hat{o}$le.

In this paper, we do not (yet) include phonons even if we acknowledge that  
their contribution   
could be relevant. However, on one hand, we wish to explore an 
electronic mechanism which {\em per se} leads to bound pairs. On the 
other hand, any mechanism in low-dimensional systems, like the 
Cuprates and nanotubes, must overcome somehow the problem of the 
repulsion between confined charges. This is an obvious difference 
compared to the traditional superconductors where pairs have hundredes 
of  Angstrom of space to delocalize. 
The notion that pairing can arise by a purely electronic mechanism, 
i.e. from purely repulsive electron-electron interactions, was put 
forth by Kohn and Luttinger long ago\cite{kohn}. They suggested that 
for large odd values of the relative angular momentum two electrons 
could stay enough far apart from each other to take advantage of the 
Friedel oscillations of the screened Coulomb potential. In our 
approach, based on a 2$d$ Hubbard model, the first-order Coulomb repulsion 
is removed by symmetry.
\begin{figure}[H]
\begin{center}
	\epsfig{figure=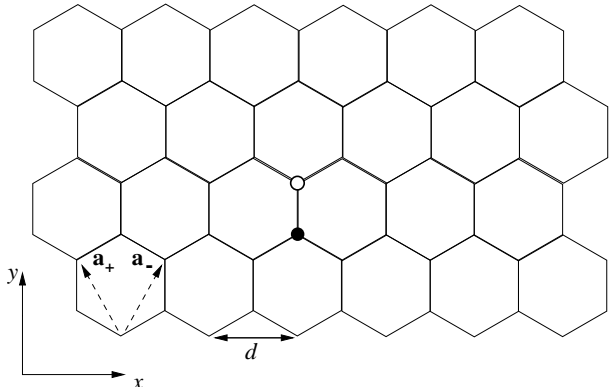,width=8cm}\caption{\footnotesize{
	Illustration of the honeycomb lattice. The sites $a$ (empty dots 
	$\circ$) and $b$ (full dots $\bullet$) constitute the 
	basis of each unit cell. We fix the origin in an $a$ site. The $(N,M)$ nanotube is 
	obtained by identifying sites separated by $N{\bf a}_{+}+M{\bf 
	a}_{-}$.
	}}
\label{honey}
\end{center}
\end{figure}

The plan of the paper is the following. 
In Section \ref{hubmod} we define the Hubbard model Hamiltonian $H$ for the armchair 
$(N,N)$ SWNT and we introduce some useful notations. In Section \ref{w=0sec} 
we use a general theorem to obtain all the two-body singlet eigenstates 
of $H$ with 
vanishing Hubbard repulsion ($W=0$ pairs). We exploit the Space Group 
symmetry of the system to get all the $W=0$ pairs with zero total momentum.  
Remarkably, we find that their wave-function vanishes if the 
particles have the same $x$ value. In Section \ref{pairmech} 
we propose a non-perturbative canonical transformation to deal with the 
effective interaction between the electrons of a $W=0$ pair added to 
the many-body ground state. Since the two extra particles cannot 
interact directly by definition of $W=0$ pair, their effective interaction
comes out from virtual electron-hole  excitation exchange with the 
Fermi sea and in principle can be attractive. 
In Section \ref{11nanotube} we consider the $(1,1)$ nanotube, with length 
$l=2d$. This system has 8 sites and is the smallest nanotube-like cluster 
showing the superconducting $W=0$ pairing. Therefore, it represents a very 
good probe to test the pairing mechanism described in Section \ref{pairmech}, 
since we can compare exact diagonalization results with the analytic ones. 
In Section \ref{superc} we apply the canonical transformation approach 
to study the $(N,N)$ nanotubes of finite length and periodic boundary 
conditions. We obtain a Cooper-like equation for the binding energy 
$-\D$ of the $W=0$ pair which is numerically solved for $2\leq N\leq 6$ and 
length $l$ up to 32$d$; the results are then extrapolated to study the 
dependence of $\D$ on the radius of the tube and on the Fermi energy in 
the limit of infinite length. Finally the conclusion are drawn in Section 
\ref{conc}.

\section{Hubbard Model on the SWNT}
\label{hubmod}
The four outer-shell electrons of each carbon atom form three 
$\sigma$ $sp_{2}$ 
bonds and a resonant $\pi$ 
bond with the remaining $p_{z}$ electron. A simple description consists 
of a tight-binding Hamiltonian where only the $p_{z}$ orbital is taken into 
account:
\begin{equation}
H_{0}=t\sum \limits_{ \langle {\bf r},{\bf r}' \rangle }\sum
 \limits_{\sigma}\left(c^{\dagger}_{{\bf r},\sigma}c_{{\bf r}',\sigma}+h.c. \right), 
\label{kin}
\end{equation}
where $c^{\dagger}_{{\bf r},\sigma}$ ($c_{{\bf r},\sigma}$) is the creation 
(annihilation) operator of an electron of spin $\s$ on the honeycomb site
${\bf r}$, the sum runs over the pairs $\langle {\bf r},{\bf r}' 
\rangle$ of nearest neighour carbon atoms and $t$ is the hopping 
parameter. The on-site Coulomb repulsion is 
\begin{equation}
W= U\sum \limits_{{\bf r}} \hat{n}_{{\bf r},\uparrow }
\hat{n}_{{\bf r},\downarrow  } ,
\label{hubrep}
\end{equation}
where $\hat{n}_{{\bf r},\s}=c^{\dagger}_{{\bf r},\sigma}c_{{\bf 
r},\sigma}$ is the number operator referred to the site ${\bf r}$ and 
to the spin $\s$.  The full Hamiltonian reads 
\begin{equation}
H=H_{0}+ W .
\label{hub}
\end{equation}
Although the the point symmetry Group of the graphite layer  is  
$D_{6h}$, the simplified Hubbard Hamiltonian in Eq.(\ref{hub})  
has $C_{6v}$ symmetry since the fermionic operators are taken to be 
even under reflection with respect to the plane; for a SWNT  
only $C_{2v}$ point symmetry remains, due to the wrapping. 
The $C_{2v}$ Group is abelian and its table of characters is
\begin{center}
    \vspace*{0.5cm}
     \begin{tabular}{|c|c|c|c|c|c|}\hline
         $C_{2v}$   &   {\bf 1}   &   $C_{2}$       &  $\sigma_{x}$   &   $\sigma_{y}$  &  Symmetry       \\ \hline
         $A_{1}$     &    1  & 1                                &  1                 &  
         1               &     $x^{2}+y^{2}$                               \\ 
         $A_{2}$     &    1  &  1                                 & -1                &  -1               &          $x/y-y/x$          \\ 
         $B_{1}$     &     1 &  -1                                  &  1               &  -1                &  $x$         \\ 
         $B_{2}$     &    1  & -1                                  &  -1                &   1             &  $y$                        
              \\ \hline
\end{tabular} 
   \vspace*{0.5cm}
\end{center}
Table I. {\footnotesize Character table ot the  C$_{2v}$ symmetry group. 
Here ${\bf 1}$ 
denotes the identity,
C$_{2}$ the 180 degrees rotation, $\sigma_{x}$  and $\sigma_{y}$ the 
reflections with respect to the
$x=0$ and $y=0$ axes respectively. The  C$_{2v}$ symmetry group is 
Abelian and hence has four one-dimensional
irreducible rapresentations (irreps) denoted by   $A_{1}$,   
$A_{2}$,   $B_{1}$,   $B_{2}$.
In the last column simple basis functions for each irrep are shown.}
\vspace*{0.5cm}

Below, we shall diagonalize the kinetic term $H_{0}$ and introduce 
some useful notation. 
Let us write the position 
$\bf R$ of the cells in terms of a pair of integers $(n,m)$,  
${\bf R}=n{\bf a}_{+}+m{\bf a}_{-}$. As illustrated in Figure 
\ref{honey}, we set the origin at an $a$ site, and therefore any 
 $\bf R$ translation will take us to another $a$ site; to get a $b$ 
 site one must translate by   ${\bf R}-(0,\frac{d}{2\sqrt{3}})$. It is convenient 
to rename the creation operators on the site ${\bf r}$ by 
distinguishing the $a$ sites from the $b$ sites:
\begin{equation}
c^{\dagger}_{{\bf r}}=
\left\{\begin{array}{ll}
c^{(a)^{\dag}}_{\bf R} & {\rm if} \;\;\;\;{\bf r}={\bf R} \\
c^{(b)^{\dag}}_{\bf R} & {\rm if} \;\;\;\;{\bf r}={\bf R}-(0,\frac{d}{2\sqrt{3}}).
\end{array}\right.
\end{equation} 
where the spin index is omitted for the sake of simplicity.
Let us introduce the Bloch creation operators  
\begin{equation}
c^{(\n)^{\dag}}_{\bf k}=\sum_{\bf R}\left[
u^{(\n)^{\ast}}({\bf k},a)c^{(a)^{\dag}}_{\bf R}+
u^{(\n)^{\ast}}({\bf k},b)c^{(b)^{\dag}}_{\bf R}\right]
e^{-i{\bf k}\cdot{\bf R}},\;\;\;\;\n=\pm
\end{equation}
with  
\begin{equation}
\left(\begin{array}{c}
u^{\pm}\left({\bf k},a\right) \\ 
u^{\pm}\left({\bf k},b\right)\end{array}\right)=
\frac{1}{\sqrt{4NL}}
\left(\begin{array}{c} 1 \\
\pm \frac{\left| A({\bf k}) 
\right|}{A({\bf k})} \end{array}\right)
\label{ufu}
\end{equation}
and
\begin{equation}
A({\bf k})=\left( 1+ 2e^{-i  \frac{d}{2} 
( \sqrt{3} k_{y}) }\cos \frac{dk_{x}}{2} 
\right).
\end{equation}

The kinetic term $H_{0}$ can be written in a diagonal form as 
\begin{equation}
H_{0}=\sum_{\n}\sum_{{\bf k},\s}
\varepsilon^{\n}\left({\bf k}\right)c^{(\n)^{\dag}}_{{\bf k},\s}
c^{(\n)}_{{\bf k},\s},
\label{kindiag}
\end{equation}
where
\begin{equation}
\varepsilon^{\pm}\left({\bf k}\right)=\pm t
\sqrt{1+4\cos^{2}\left(\frac{k_{x}d}{2}\right)+
4\cos\left(\frac{k_{x}d}{2}\right)\cos\left(\frac{k_{y}\sqrt{3}d}{2}\right)},
\end{equation}
are the bonding (-) and antibonding (+) bands.

The Hubbard interaction becomes 
\begin{equation}
W=2NL\sum_{{\bf k}_{1},{\bf k}_{2},{\bf k}_{3},{\bf k}_{4}}
\sum_{\n_{1},\n_{2},\n_{3},\n_{4}}
U_{\n_{1},\n_{2},\n_{3},\n_{4}}({\bf k}_{1},{\bf k}_{2},{\bf k}_{3},{\bf k}_{4})
\;
c^{(\n_{1})^{\dag}}_{{\bf k}_{1},\ua}c^{(\n_{2})^{\dag}}_{{\bf k}_{2},\da}
c^{(\n_{3})}_{{\bf k}_{3},\da}c^{(\n_{4})}_{{\bf k}_{4},\ua},
\label{wtrans}
\end{equation}
where
\begin{equation}
U_{\n_{1},\n_{2},\n_{3},\n_{4}}({\bf k}_{1},{\bf k}_{2},{\bf k}_{3},{\bf k}_{4})=
U\sum_{\z=a,b}
u^{(\n_{1})^{\ast}}({\bf k}_{1},\z)
u^{(\n_{2})^{\ast}}({\bf k}_{2},\z)
u^{(\n_{3})}({\bf k}_{3},\z)
u^{(\n_{4})^{\ast}}({\bf k}_{4},\z)
\d_{\bf G}({\bf k}_{1}+{\bf k}_{2}-{\bf k}_{3}-{\bf k}_{4})
\label{vertex}
\end{equation}
and $\d_{\bf G}({\bf k})$ is 1 if ${\bf k}$ is a reciprocal lattice vector 
${\bf G}$ and zero otherwise; in the $(N,N)$ nanotubes 
${\bf G}=n{\bf G}_{+}+m {\bf G}_{-}$, with $n$ and $m$ 
integers, and ${\bf G}_{\pm}=
\frac{2\p}{d}(1,\pm \frac{1}{\sqrt{3}})$.
For the $(N,N)$ nanotubes of length $l=Ld$ and periodic boundary 
conditions along the $\hat x$ direction the ${\bf k}$-vectors are 
quantized as 
\begin{equation}
k_{x}=\frac{2\p}{Ld}m_{x},\;\;\;\;m_{x}=0,1,\ldots,L-1 \; ;
\;\;\;\;\;\;\;\;
k_{y}=\frac{2\p}{\sqrt{3}Nd}m_{y},\;\;\;\;m_{y}=0,1,\ldots,2N-1 \;.
\label{kquan}
\end{equation}
For ${\bf k}=(\pm \frac{2}{3d} \pi ,\frac{2}{\sqrt{3}d} \pi )\equiv 
{\bf k}_{\pm}$ the bonding and antibonding bands touch each other and 
$\varepsilon^{\pm} \left({\bf k}_{\pm}\right)=0$  linearly.

In the next Section we show that the Hamiltonian in Eq.(\ref{hub}) 
admits two-body singlet eigenstates with no
double occupancy on the honeycomb sites and we shall refer to them 
as $W=0$ pairs. $W=0$ pairs are therefore  
eigenstates of the kinetic energy operator in Eq.(\ref{kindiag}) 
and of the Hubbard repulsion $W$ of Eq.(\ref{wtrans}) with vanishing 
eigenvalue of the latter. The particles forming a $W=0$ pair have no 
direct interaction and are the main candidates to achieve bound 
states in purely repulsive Hubbard 
models\cite{EPJB1999}\cite{EPJB2001}\cite{cibal}.
Incidentally, we note that such states are involved in the 
antiferromagnetic ground state of Hubbard and related models at half 
filling\cite{jop2001}\cite{ssc2001}\cite{jop2002}\cite{tobepub}.

\section{$W=0$ Pairs in the armchair $(N,N)$ SWNT}
\label{w=0sec}

We obtained\cite{EPJB2001}\cite{IJMPB2000} a powerful and elegant criterion to get 
{\em all} the $W=0$ pairs. We can do that in terms of the Optimal Group  ${\cal G}$ 
of the Hamiltonian, that we define as a symmetry Group which is 
big enough to justify the degeneracy of 
the single particle energy levels.  By definition, every one-body eigenstate of $H$ can be 
classified as belonging to one of the irreducible representations (irreps) of ${\cal G}$. We may 
say that an irrep $\eta$ is represented in the one-body spectrum of $H$ if 
at least one of the one-body levels belongs to $\eta$. Let
${\cal E}$ be the set of the irreps of ${\cal G}$ which are 
represented in the one-body spectrum of $H$. Let $|\q\rangle$ be a two-body eigenstate of 
the kinetic energy $ H_{0}$ with spin $ S_{z}=0$. Then, it holds the 

{\it W=0 Theorem}:  
\begin{equation}
\eta \notin  {\cal E} \Leftrightarrow W P^{(\eta)}|\q\rangle=0 \;
\label{theo}
\end{equation}
where $P^{(\eta)}$ is the 
projection operator on the irrep $\eta$. In other terms, any nonvanishing projection 
of $|\q\rangle$ on an irrep
{\em not} contained in ${\cal E}$, is an eigenstate of $ H_{0}$ 
with no double occupancy. The singlet component of this state is a  
$W=0$ pair. Conversely, any pair belonging to an irrep represented in the one-body
spectrum must have positive $W$ expectation value. 

The complete characterization of the symmetry of $W=0$ pairs requires 
the knowledge of the Optimal Group ${\cal G}$. A partial use of the 
theorem is possible if one  does not know 
${\cal G}$ but knows a subgroup.
It is then still granted that any pair belonging to an irrep of the 
subgroup not represented in the spectrum has the $W=0$ property.  On the other 
hand, accidental degeneracies occur with a subgroup of the Optimal 
Group, because by mixing degenerate pairs belonging to irreps represented 
in the spectrum one can find $W=0$ pairs also there. This is  
illustrated by the example reported in Section \ref{11nanotube}. 

Below, we shall apply the {\em $W=0$ Theorem} in the $(N,N)$ SWNT to obtain 
$W=0$ pairs of zero total-momentum. Let $|0\ket$ denote the electron 
vacuum. Exploiting the invariance of the 
Hamiltonian under translations and $C_{2v}$-operations the determinantal 
state $c^{\dag}_{{\bf k},\ua}c^{\dag}_{-{\bf k},\da}|0\ket$ yields 
nothing if projected onto an irrep of the Space Group with non-zero 
momentum. Direct inspection of the Bloch functions in Eq.(\ref{ufu}) shows that the ${\bf k}=0$ irreps represented in the one-body spectrum 
are $B_{2}$ in the bonding band and $A_{1}$ in the antibonding band; 
hence, $W=0$ pairs may be obtained by projecting 
onto $A_{2}$ and $B_{1}$. Let us consider the two-electron singlet 
state of vanishing momentum:
\begin{eqnarray}
\q_{\z_{1},\z_{2}}\left({\bf k}, {\bf R}_{1} , {\bf R}_{2} \right) 
&\equiv& 
\frac{1}{\sqrt{2}}
\bra 0|c_{R_{1},\ua}^{(\z_{1})}c_{R_{2},\da}^{(\z_{2})}
[c^{\dag}_{{\bf k},\ua}c^{\dag}_{-{\bf k},\da}+
c^{\dag}_{-{\bf k},\ua}c^{\dag}_{{\bf k},\da}]\;|0\ket =
\nonumber \\ &=&
\frac{1}{\sqrt{2}}\left[ u^{\ast}\left({\bf k},\z_{1}\right) 
u^{\ast}\left(-{\bf k},\z_{2}\right) e ^{i{\bf k}\cdot ({\bf R}_{1}-
{\bf R}_{2}  )}+ u^{\ast}\left({\bf k},\z_{2}\right) u^{\ast}
\left(-{\bf k},\z_{1}\right) e ^{-i{\bf k}\cdot ({\bf R}_{1}-
{\bf R}_{2}  )}  \right] \chi_{0}, 
\end{eqnarray}
where $\chi_{0}$ is a singlet spin function and for 
$i=1,2$ $\z_{i}=a,b$. The projection onto the irrep $\eta$ of $C_{2v}$ yields  
\begin{equation}
\q^{[\eta]}_{\z_{1},\z_{2}}\left({\bf k}, {\bf R}_{1} , {\bf R}_{2} \right) =  
\frac{1}{2} \sum_{\hat{O} \in C_{2v} } \chi ^{ (\eta) }(\hat{O}) 
\q_{\z_{1},\z_{2}}\left(\hat{O}{\bf k}, {\bf R}_{1} , {\bf R}_{2} \right)  ,
\end{equation}
where $ \chi ^{ (\eta) }(\hat{O})$ is the character in $\eta$ of the  operation 
$\hat{O}$ of $C_{2v}$. While 
 $\q^{[B_{1}]}=0$, after some algebra one finds  
\begin{eqnarray}
\q^{[A_{2}]}_{\z_{1},\z_{2}}\left({\bf k}, {\bf R}_{1} , {\bf R}_{2} 
\right) &=& 
\sin \left( k_{x}(X_{1}-X_{2})\right) \times 
\nonumber \\ &\times& \frac{1}{\sqrt{2}} 
\left[u^{\ast}\left({\bf k},\z_{1}\right)
u^{\ast}\left(-{\bf k},\z_{2}\right) 
e^{ik_{y}(Y_{1}-Y_{2})}-
u^{\ast}\left({\bf k},\z_{2}\right) 
u^{\ast}\left(-{\bf k},\z_{1}\right)
e^{-ik_{y}(Y_{1}-Y_{2})}\right]\, \chi_{0}, 
\end{eqnarray}
where ${\bf R}_{i,j}=(X_{i,j},Y_{i,j})$. We can verify by direct 
inspection that $\q^{[A_{2}]}_{\z_{1},\z_{2}}({\bf k},{\bf R}_{1},{\bf R}_{2})$ 
vanishes for $X_{1}=X_{2}$, that is the two-body singlet 
wavefunction vanishes if the particles lie on the same annulus of 
the $(N,N)$ tube. As a consequence 
$\q^{[A_{2}]}_{\z_{1},\z_{2}}({\bf k},{\bf R}_{1},{\bf R}_{2})$ is 
an eigenstate of the kinetic energy $H_{0}$ [with eigenvalue 
$2\ve({\bf k})$)] and of the on-site Hubbard repulsion $W$ with 
vanishing eigenvalue of the latter, that is 
$\q^{[A_{2}]}_{\z_{1},\z_{2}}({\bf k},{\bf R}_{1},{\bf R}_{2})$ is a 
$W=0$ pair. 

$W=0$ pairs of non-vanishing total-momentum may be obtained in a 
similar way; however, in this preliminary work we concentrate on 
pairs of vanishing total momentum. 

\section{Canonical transformation approach to the pairing mechanism}
\label{pairmech}

In this Section we intend to study the 
effective interaction among the 
electrons of a $W=0$ pair added to the $n$-body interacting ground 
state $|\Psi_{0}(n) \rangle$. Since the two extra particles cannot 
interact directly by definition of $W=0$ pair, their effective interaction
comes out from virtual electron-hole  excitation exchange with the 
Fermi sea and in principle can be attractive. 

Many configurations contribute to the interacting $(n+2)$-body 
ground state $|\Psi_{0}(n+2) \rangle$
and we need a complete set $\cal{S}$ to expand it exactly; as 
long as it is complete, however, we can design 
$\cal{S}$ as we please. We can take the non-interacting  $n$-body 
Fermi {\em sphere} $|\F_{0}(n)\rangle$ as our vacuum
and build the complete set  in terms of excitations over it. 
In the subspace with vanishing spin $z$ component, the simplest states 
that enter the configuration mixing are those obtained from 
$|\F_{0}(n)\rangle$ by creating two extra electrons over it; we 
denote with $|m\ket$ these states.  
Similarly, along with the  pair $m$ states, we introduce
the  4-body $\alpha $ states, obtained from $|\Phi_{0}(n)\ket$ by creating $2$ 
electrons and 1 electron-hole (e-h) pair.   
Then  $\mathcal{S}$ includes the  6-body $\beta$ states  having $2$ electrons 
and 2 e-h pairs, and so on. 
We are using Greek indices for the configurations containing the electron-hole 
pairs, which here are playing largely the same r$\hat{o}$le as 
phonons in the Cooper theory. By means of the complet set $\mathcal{S}$ 
we now expand the interacting ground state 
\begin{equation}
|\Psi _{0}(n+2)\ket={\sum_{m}}a_{m}|m\ket+{\sum_{\alpha }}
a_{\alpha }|\alpha \ket+{\sum_{\beta }}a_{\beta }
|\beta \ket+....  
\label{lungo}
\end{equation}
and set up the Schr\"{o}dinger equation
\begin{equation}
H|\Psi_{0}(n+2)\ket=E(n+2)|\Psi _{0}(n+2)\ket.
\label{seq}
\end{equation}
We stress that Eq.(\ref{lungo}) is configuration interaction, 
{\em not a perturbative expansion}.
When  the number $n$  of electrons in the  system is such that 
$|\Phi_{0}(n)\ket$ is a single non-degenerate determinant 
(the Fermi surface is totally filled),  we can easily and 
unambiguously define and calculate  the effective interaction
between the
two extra electrons since the expansion in Eq.(\ref{lungo}) for the 
interacting ground state is unique: this is done  by a 
canonical transformation\cite{EPJB1999},\cite{cbs1},\cite{cbs3} from the 
many-body Hamiltonian of Eq.(\ref{hub}). We
consider the effects of the operators $H_{0}$ and $W$ 
on the terms of $|\Psi _{0}(n+2)\ket$. Choosing the $m$, $\a$, $\b$, 
$\ldots$ states to be eigenstates of the kinetic energy $H_{0}$ we 
have 
\begin{equation}
H_{0}|m\ket=E_{m}|m\ket,  \;\;\;\;
H_{0}|\a\ket=E_{\a}|\a\ket,  \;\;\;\;
H_{0}|\b\ket=E_{\b}|\b\ket,  \;\;\;\;\ldots.
\label{h0effect}
\end{equation}
Since $W$ can create or destroy up to 2 e-h pairs, its action on an 
$m$ state yields
\begin{equation}
W|m\ket={\sum_{m^{\prime }}}
W_{m^{\prime },m}|m^{\prime }\ket+
\sum_{\alpha}W_{\alpha,m}|\alpha\ket +
\sum_{\beta}W_{\beta ,m}|\beta\ket .  
\label{weffect}
\end{equation}
The action of $W$ on the $\a$ states yields 
\begin{equation}
W|\alpha\ket={\sum_{m}}W_{m,\alpha }|m\ket +
\sum_{\alpha^{\prime }}W_{\alpha ^{\prime },\alpha }|\alpha^{\prime }\ket+
\sum_{\beta} W_{\beta ,\alpha}|\beta\ket+
\sum_{\g}W_{\g\a}|\g\ket ,
\label{weffect2}
\end{equation}
where scattering between 4-body states is allowed by the second term, 
and so on. In this way we obtain an algebraic system for the coefficients 
of the configuration interaction of Eq.(\ref{lungo}). However to test the instability
of the Fermi liquid towards pairing it is sufficient to study the amplitudes 
$a_{m}$ of the $m$ states. In the weak coupling limit this can be done 
by 
truncating the expansion in Eq.(\ref{lungo}) to the $\a$ states because, 
as we have shown\cite{cbs1}, the inclusion of the $\b,\;\g,\ldots$ 
states produces a $E$-dependent renormalization of the matrix elements 
of higher order in $W$, leaving the structure of the equations unaltered.

By taking a linear combination of the $\a$ states in such a way that
\begin{equation}(H_{0}+W)_{\a,\a'}=
\d_{\a\a'}E'_{\a}
\end{equation}
the algebraic system reduces to
\begin{equation}
\left[ E_{m}-E(n+2)\right]
a_{m}+{\sum_{m^{\prime }}}a_{m^{\prime}}
W_{m,m^{\prime }}
+{\sum_{\alpha }}a_{\alpha }W_{m,\alpha }=0 
\label{reson1}
\end{equation}
\begin{equation}
\left[ E'_{\alpha }-E(n+2)\right] a_{\alpha }+
{\sum_{m^{\prime }}}a_{m^{\prime}}W_{\alpha ,m^{\prime }}=0.  
\label{reson2}
\end{equation}
Solving for $a_{\alpha }$ and 
substituting in Eq.(\ref{reson1}) we exactly decouple the
4-body states as well, ending up with an equation for the dressed 
pair $|\q\rangle=\sum_{m}a_{m}|m\rangle$. 
The effective Schr\"{o}dinger equation for the pair reads
\begin{equation}
\left(H_{0}+W+S[E]\right) |\q\ket \equiv H_{\rm pair} |\q\ket =E|\q\ket  
\label{can1}
\end{equation}
where
\begin{equation}
(S[E])_{m,m'}=-\sum_{\a}\frac{W_{m,\a}W_{\a,m'}}{E'_{\a}-E}.
\label{fscat}
\end{equation}
is the scattering operator. 
The matrix elements $W_{m,m'}$ in Eq.(\ref{can1}) may be written as the sum of two terms 
representing the direct interaction $W^{(d)}_{m,m'}$ among the 
particles forming the pair and the first-order self-energy $W_{m}$:
\begin{equation}
W_{m,m'}=W^{(d)}_{m,m'}+\d_{m,m'}W_{m}.
\label{selfene}
\end{equation}
Analogously in $S[E]$ we may recognize two different 
contributions; one is the 
true effective interaction $W_{\rm eff}$ between the 
electrons of the $m$ states, while the other one is the forward scattering 
term $F$ 
\begin{equation}
S_{m,m'}=(W_{\rm eff})_{m,m'}+F_{m}\d_{m,m'} \,.
\end{equation}
The first-order self-energy and the forward scattering term are diagonal in the 
indices $m$ and $m'$.  $W_{m}$ and $F_{m}$ renormalize the non-interacting 
energy $E_{m}$ of the $m$ states: 
\begin{equation}
    E_{m}\ra E^{(R)}_{m}=E_{m}+W_{m}+F_{m}.
\label{eneren}
\end{equation}
Eq.(\ref{can1}) is of the form of a Schr\"odinger equation with eigenvalue 
$E(n+2)$ for the added pair with the interaction  $W^{(d)}+W_{\rm eff}$. 
Here the $W=0$ pairs are special because $W^{(d)}$ vanishes. 
We interpret $a_{m}$ as the wave function of the dressed pair, 
which is acted upon by an effective Hamiltonian $H_{\rm pair}$. This 
way of looking at Eq.(\ref{can1}) is perfectly consistent, despite the 
presence of  the many-body eigenvalue $E(n+2)$.
Indeed, if the interaction is attractive and produces bound states the 
spectrum of Eq.(\ref{can1}) contains discrete states below the threshold of 
the continuum (two-electron Fermi energy).
This is a clear-cut criterion for pairing, which is exact in principle. 
The threshold is given by 
\begin{equation}
E^{(R)}_{F}\equiv \min_{\{m\}}[ E^{(R)}_{m}(E)],
\label{min}
\end{equation}
which contains all the pairwise interactions except those between 
the particles in the pair; it must be determined once 
Eq.(\ref{can1}) has been solved (since $F$ depends on the solution). 
The ground state energy $E$ may be conveniently written as $E^{(R)}_{F}+\D$. 
$\D<0$ indicates a Cooper-like instability of the normal Fermi liquid 
and its magnitude  
represents the binding energy of the pair. 
 
We emphasize the fact  that in  principle the canonical transformation is exact
because in this way our framework does not require $U/t$ to be small.
The next problem is how to find a practical estimate of the 
renormalized Fermi energy. 
In the numerical calculations, some approximation is needed. In 
Section \ref{11nanotube} and \ref{superc}, 
we shall compute the bare quantities; that is, we shall neglect the 6-body 
and higher excitations in the calculation of $W_{\rm eff}$ and $F$.  
This is a resonable approximation if we compute small corrections to a 
Fermi liquid background and the exact numerical results in the $(1,1)$ nanotube
suggest that this is the case, see Section \ref{11nanotube}.   

We want to stress that once the expansion of $|\Q_{0}(n+2)\ket$ 
in Eq.(\ref{lungo}) is truncated as specified above we do not need to 
construct  a   good approximation of the interacting ground 
state wave function in order to get the  $a_{m}$ amplitudes at weak 
coupling; in this 
way we obtain information about pairing. 

\section{The $(1,1)$ nanotube: exact diagonalization and pairing mechanism}
\label{11nanotube}

There is evidence from cluster calculations  that pairing can 
arise in the repulsive Hubbard model\cite{cibal}\cite{citro}\cite{fettes}. 
In this Section we shall consider the $(1,1)$ nanotube, with  
$l=2d$ and periodic boundary conditions, see Fig(\ref{tubo11}.$a$). This system 
has 8 sites and is the smallest nanotube-like cluster showing the superconducting 
$W=0$ pairing. Therefore, it represents a very good probe to test the pairing 
mechanism shown in Section \ref{pairmech}, since we can compare exact diagonalization 
results with the analytic approximations of the canonical 
transformation. 

We define, following Refs.\cite{scalah}\cite{bal},
\begin{equation}
{\tilde \Delta}(n+2)=E(n+2)+E(n)-2E(n+1). 
\label{delta}
\end{equation}
where $E(n)$ is the ground state energy with $n$ electrons (referenced 
to the electron vacuum).
$|{\tilde \Delta}(n+2)|$ is one definition of the pairing energy. 
This definition is simple, but requires computing the eigenvalues 
with great accuracy, and has several drawbacks. It says nothing about 
the dynamics which leads to pairing. Moreover, generally a negative 
${\tilde \Delta}$ does not unambiguously imply
pairing, and further problems arise since the 
above definition depends on the comparison of systems with different 
$n$. 

However,  in several 
studies of $W=0$ pairing in finite systems when it was possible to 
compute  ${\tilde \Delta}$ by exact diagonalization we pointed 
out\cite{EPJB2001}\cite{cibal}\cite{cbs3}  
that at least at weak coupling  it agrees well with   $\D$ as obtained by the 
canonical transformation. This supports the application of Eq.(\ref{delta}). 

Below we perform a group-theoretical analysis and obtain $W=0$ pairs 
by exploiting the {\em  $W=0$ Theorem}. Next, we compute the interacting ground 
state energy  with 2, 3 and 4 electrons by exact numerical 
diagonalization in order to get ${\tilde \Delta}(4)$. 
Finally the canonical transformation is applied to evaluate $\D(4)$ 
which will be compared with ${\tilde \Delta}(4)$.

\subsection{Symmetry Properties and $W=0$ Pairs}
\label{spin11}

The First Brillouin Zone (FBZ) consists of 4 points and since there are 
two atoms in the unit cell, two bands result. The symmetry properties of this 
system are intriguing: it is not only invariant under the operations of the 
Space Group (translations and $C_{2v}$-operations), but also under the 
dynamical operation $d$ shown in Fig.(\ref{tubo11}.$b$). 
The {\em dynamical} operation $d$ is reminiscent of a similar symmetry 
which must be taken into account to understand the degeneracies in 
the $4 \times 4$ Hubbard model \cite{EPJB2001}. 
\begin{figure}[H]
\begin{center}
	\epsfig{figure=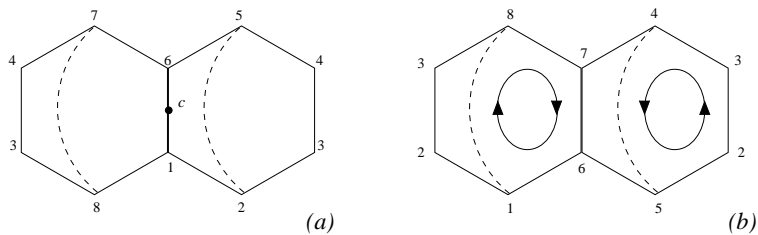,width=10cm}\caption{\footnotesize{
	($a$) The Hubbard model of the (1,1) nanotube; the dashed lines 
	denote hopping interactions due to the wrapping.
	($b$) Illustration of the dynamical symmetry $d$; the sites 1, 8, 7, 
	6 undergo a clockwise rotation while the sites 3, 2, 5, 4 a 
	counterclockwise rotation. One can see by direct inspection that 
	the nearest neighbours of each site are the same of Fig.(\ref{tubo11}.$a$). 
	}}
\label{tubo11}
\end{center}
\end{figure}

Including  $d$ and closing the multiplication table
we obtain a symmetry Group ${\cal G}$  with 48 elements in 
10 classes as shown in Table II.
\begin{center}
    \vspace*{0.5cm}
     \begin{tabular}{|c|c|c|c|c|c|c|c|c|c|}
	\hline
	${\cal C}_{1}$&${\cal C}_{2}$&${\cal C}_{3}$&${\cal C}_{4}$&${\cal C}_{5}$&
	${\cal C}_{6}$&${\cal C}_{7}$&${\cal C}_{8}$&${\cal C}_{9}$&${\cal C}_{10}$\\
	\hline
	${\bf 1}$&$C_{2}$&$\sigma_{x}$&$\sigma_{y}$&
	$\sigma_{x}t_{1,0}^{2}$&$t_{1,0}^{2}$&$t_{1,0}$
	&$d$&$C_{2}d$&$\sigma_{y}d$\\
	\hline
\end{tabular}
   \vspace*{0.5cm}
\end{center}
Table II. {\footnotesize  Top row: symbols of the 10 classes ${\cal C}_{i}$ of ${\cal 
G}$; bottom row: one typical operation for each of the 
classes; the others can be obtained by conjugation. 
The operations are: the identity ${\bf 1}$, the translation $t_{n,m}$ of 
$n$ steps along ${\bf a}_{+}$ and $m$ along ${\bf{a}_{-}}$; $d$ is the 
{\em dynamical} symmetry. The other operations $C_{2}$, $C_{4}$, 
$\sigma_{x}$, $\sigma_{y}$ are those of the Group of 
the rectangle and are referenced to the centre $c$ of 
Fig(\ref{tubo11}.$a$).} 
\vspace*{0.5cm}

In Table III we report the character table of the full symmetry Group  
$\cal{G}$ of the $(1,1)$ nanotube of length $l=2d$.

\begin{center}
\begin{tabular}{|l|l|l|l|l|l|l|l|l|l|l|}
\hline
$\cal{G}$ & ${\cal C}_{1}$ & ${\cal C}_{2}$ & ${\cal C}_{3}$ & 
${\cal C}_{4}$ &  ${\cal C}_{5}$& ${\cal C}_{6}$ &  
${\cal C}_{7}$& ${\cal C}_{8}$ & ${\cal C}_{9}$ &  ${\cal C}_{10}$\\
\hline
 $A_{1}$&  1  &  1  &  1  &  1  &  1  &  1  &  1  &  1  &  1  &  1  \\
\hline
$A_{2}$ &  1  &  1  & -1  & -1  & -1  &  1  & -1  &  1  &  1  & -1  \\
\hline 
 $B_{1}$&  1  & -1  &  1  & -1  &  1  &  1  & -1  & -1  &  1  &  1 \\
\hline
$B_{2}$ &  1  & -1  & -1  &  1  & -1  &  1  &  1  & -1  &  1  & -1 \\
\hline
$E_{1}$ &  2  &  0  &  2  &  0  &  2  &  2  &  0  &  0  & -1  & -1 \\
\hline
 $E_{2}$&  2  &  0  & -2  &  0  & -2  &  2  &  0  &  0  & -1  &  1 \\
\hline
$T_{1}$ &  3  &  1  & -1  &  1  &  3  & -1  & -1  & -1  &  0  &  0 \\
\hline
$T_{2}$ &  3  & -1  & -1  & -1  &  3  & -1  &  1  &  1  &  0  &  0 \\
\hline
$T_{3}$ &  3  & -1  &  1  &  1  & -3  & -1  & -1  &  1  &  0  &  0 \\
\hline
$T_{4}$ &  3  &  1  &  1  & -1  & -3  & -1  &  1  & -1  &  0  &  0 \\
\hline
\end{tabular}
\end{center}
Table III. {\footnotesize Character table ot the  Optimal Group $\cal{G}$. 
 The irreps $A_{1}$, $A_{2}$, $B_{1}$ and $B_{2}$ reduce to the corresponding 
ones of the subgroup $C_{2v}$ if  ${\cal G}$ is broken.}
\vspace*{0.5cm}

The one-body eigenenergies for $t=1$ eV are shown in Table IV 
together with the irreps of the associated eigenvectors. 
From Table II and III we see that $\cal{G}$ is an Optimal Group as 
defined above.
\begin{center}
\begin{tabular}{lllll}
Quasi-momentum\quad\quad & 
Energy ($\n=-$)\quad\quad & 
Irrep\quad\quad & 
Energy ($\n=+$)\quad\quad & 
Irrep\quad\quad \\
${\bf k}_{1}=(0,0)$ & $\quad$ -3 &$B_{2}$& $\quad$ 3 & $A_{1}$ \\
$ {\bf k}_{2}=(0, \frac{2 \pi}{ \sqrt{3} } ) $ & $\quad$ -1 &$T_{1}$ & 
$\quad$ 1 & $T_{3}$ \\
${\bf k}_{3}=(\pi,0)$ & $\quad$ -1 & $T_{1}$ &$\quad$ 1 & $T_{3}$ \\
${\bf k}_{4}= (- \pi, 0 ) $ & $\quad$ -1 & $T_{1}$ &  $\quad$ 1 & $T_{3}$
\end{tabular}
\end{center}
\begin{center}
Table IV. {\footnotesize One-body spectrum for $t=1$ eV. The energies 
are in eV.}
\end{center}
\vspace*{0.5cm}

In Section \ref{w=0sec} we have shown how to get all the $W=0$ 
pairs of vanishing momentum and belonging to the irrep $A_{2}$ of 
$C_{2v}$. However, the $(1,1)$ nanotube is too small to 
achieve this kind of $W=0$ pair-states and the projection on the irrep 
$A_{2}$ of $c^{\dag}_{{\bf k},\ua}c^{\dag}_{-{\bf k},\da}|0\ket$ is 
identically zero. Nevertheless, this cluster admits $W=0$-pair 
solutions of different type and they may be obtained by applying
the {\em  $W=0$ Theorem}\cite{IJMPB2000}, which contains a very general 
prescription to determine {\it all} the $W=0$ pairs, see Section 
\ref{w=0sec}.

In the $(1,1)$ nanotube $W=0$ pairs formed by particles of the same kinetic energy may be 
obtained in the 3-fold degenerate one-body levels with energies $\pm$1 eV 
by projecting $c^{\dag}_{{\bf k}_{i},\ua}c^{\dag}_{{\bf k}_{j},\da}|0\ket$ 
with $i,j=2,3,4$, onto the irreps which are not represented in the one-body spectrum. 
In this way we find two singlets
\begin{eqnarray}
| \psi_{1}^{^{1} E _{1}} \rangle = \frac{1}{\sqrt{2}}  
(c^{\dagger}_{{\bf k}_{3},\uparrow} c^{\dagger}_{{\bf k}_{3},\downarrow} 
- c^{\dagger}_{{\bf k}_{4},\uparrow} 
c^{\dagger}_{{\bf k}_{4},\downarrow} ) | 0 \rangle 
\label{pair1} \\
| \psi_{2}^{^{1} E _{1}} \rangle = 
\frac{1}{\sqrt{6}}  (2\,c^{\dagger}_{{\bf k}_{2},\uparrow} c^{\dagger}_{{\bf k}_{2},\downarrow} 
- c^{\dagger}_{{\bf k}_{3},\uparrow} 
c^{\dagger}_{{\bf k}_{4},\downarrow}- c^{\dagger}_{{\bf k}_{4},\uparrow} 
c^{\dagger}_{{\bf k}_{3},\downarrow}) | 0 \rangle 
\label{pair2}
\end{eqnarray}
and three triplets 
\begin{eqnarray}
| \psi_{1}^{^{3} T _{2}} \rangle = \frac{1}{\sqrt{2}}  
(c^{\dagger}_{{\bf k}_{3},\uparrow} c^{\dagger}_{{\bf k}_{4},\downarrow} 
- c^{\dagger}_{{\bf k}_{4},\uparrow} 
c^{\dagger}_{{\bf k}_{3},\downarrow} ) | 0 \rangle \label{trip1}\\
 | \psi_{2}^{^{3} T _{2}} \rangle = 
\ \frac{1}{\sqrt{2}}  (c^{\dagger}_{{\bf k}_{3},\uparrow} c^{\dagger}_{{\bf k}_{2},\downarrow} 
- c^{\dagger}_{{\bf k}_{2},\uparrow} 
c^{\dagger}_{{\bf k}_{3},\downarrow} ) | 0 \rangle \\  | \psi_{3}^{^{3} T _{2}} \rangle = 
\ \frac{1}{\sqrt{2}}  (c^{\dagger}_{{\bf k}_{4},\uparrow} 
c^{\dagger}_{{\bf k}_{2},\downarrow} 
- c^{\dagger}_{{\bf k}_{2},\uparrow} 
c^{\dagger}_{{\bf k}_{4},\downarrow} ) | 0 \rangle 
\label{triptre}
\end{eqnarray}
while all the other projections yield nothing.  

In the above expressions the Bloch states $c^{\dagger}_{{\bf k}_{i} ,\sigma} 
|0 \rangle $ can be taken both in the bonding and the antibonding bands ($\ve =\mp 1$ 
eV respectively). In conclusions we found that the $(1,1)$ nanotube
with 8 sites has two  $W=0$ pairs in each band. As already observed, such pairs 
have a non-vanishing total-momentum, contrarily to the $W=0$ pairs of 
Section \ref{w=0sec}. 

\subsection{Exact diagonalization data}

We have performed the numerical diagonalization of the Hamiltonian in 
Eq.(\ref{hub}) for the $(1,1)$ nanotube of length $2d$ and periodic 
boundary conditions, filled with $2,\;3$ and 4 electrons. 
The interesting case arises when the number of electrons is $n+2=4$, since 
the non-interacting ground state $|\F_{0}(2)\ket$ is a non-degenerate 
singlet with vanishing momentum containing two electrons in 
the lowest energy level. Therefore, the two added electrons may form a $W=0$ pair 
according to the result of Section \ref{spin11}. 

We found that the interacting ground state $|\Q_{0}(2)\ket$ with 2 
electrons is a totally symmetric singlet with vanishing momentum 
for any value of the ratio $U/t$. On the other hand, $|\Q_{0}(3)\ket$
is three-fold degenerate in the sector $S_{z}=\frac{1}{2}$, with momenta 
$(0,\frac{2 \pi}{\sqrt{3}})$, $(\pi,0)$ and $(-\pi,0)$, for $U/t<10^{5}$. 
It belongs to the irrep $T_{1}$ in the weak coupling regime. 
With four electrons (784 configurations) the ground state is a doubly degenerate
singlet, with 
momenta $(0,0)$ and $(0,\frac{2\pi}{\sqrt{3}})$, and belongs to  
the irrep $^{1}E_{1}$. This is the symmetry of the $W=0$ pair; our 
approach predicts the correct symmetry of the ground state.
The first excited state with 4 electrons is a three-fold 
degenerate triplet with momenta  $(0,0)$, $(\pi,0)$ 
and $(-\pi,0)$ and belongs to $T_{2}$. We show below that we are able 
to predict the symmetry of this state as well. We found 
that there is no level-crossings up to $U/t<10^{5}$. 

In order to study the pairing between the two added electrons, 
we must compute the quantity ${\tilde \Delta}(4)=E(4)+E(2)-2\,E(3)$, which 
is related to the effective interaction according to the discussion 
made at the beginning of Section \ref{11nanotube}. 
${\tilde \Delta}(4)$ has been computed in a large range of $U/t$ values, 
and its trend is shown in Fig.(\ref{delta11}). 
\begin{figure}[H]
\begin{center}
	\epsfig{figure=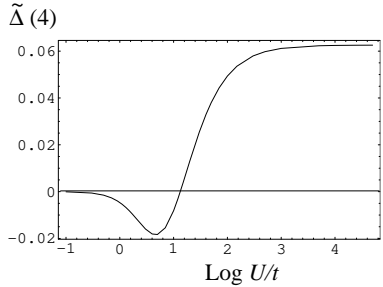,width=5cm}\caption{\footnotesize{
	Trend of ${\tilde \Delta}(4)$ versus ${\mathrm Log}\; U/t$ for 
	$U/t$ in the range $1\div 10^{5}$, ${\tilde \Delta}(4)$ is in eV. 
	}}
\label{delta11}
\end{center}
\end{figure}
For arbitrary small values of $U/t$, ${\tilde \Delta}(4)$ is negative and decreases up to 
a characteristic value of $U/t \sim 4 \div 5$,
where a minimum is reached; at the minimum ${\tilde \Delta}(4) \sim -0.018$eV. 
Thus, as in the model Cu-O clusters \cite{cibal}, ${\tilde \Delta}(4)$ 
is a new energy scale of the system, widely different from any of the 
input parameters. 
For larger $U/t$, ${\tilde \Delta}(4)$ increases until
${\tilde \Delta}(4)=0$ for $ U/t \sim 14$. As far as $U/t\to \infty$, ${\tilde \Delta}(4)$ 
increases monotonically up to the asymptotic
positive value of $\sim 0.063$ eV. We emphasize that ${\tilde \Delta}(4)$ becomes 
positive for large values of $U/t$ and hence
pairing disappears in the strong coupling regime. Therefore, the above 
pairing mechanism cannot be related to the
ones considered  within the framework of the $t-J$-like models, where 
an infinite $U$ is required to forbid double 
occupation on the same site and pairing is achieved by means of residual 
attractive interactions. 

\subsection{Analytical Canonical Tranformation: Pairing Mechanism}

In this Section we study the $(1,1)$ nanotube with 8 sites by 
implementing the canonical transformation  
described in Section \ref{pairmech} but truncated to the $\a$-states.  
We shall compare the analytic results with the numerical data obtained previously. 
The non-interacting Fermi {\em sphere} $|\F_{0}(2) \rangle = 
c^{\dagger}_{k_{1}\uparrow} c^{\dagger}_{k_{1}\downarrow}|0\rangle$ 
has two electrons in the lowest level. Since the one-body levels are 
widely separated the intra-shell interaction is much more important than 
the inter-shell one. Therefore, 
we consider only the $|m\rangle$ states having the added pair of electrons in 
the lowest unoccupied level of energy -1 eV, neglecting the 
higher-energy orbitals. We recall that in the three-fold degenerate 
level with energy -1 eV we can write 5 two-body states (2 singlets and 3 
triplets) without double occupancy, see Eqs.(\ref{pair1}-\ref{triptre}). 
Hence, we may further reduce the set of the $m$ states in the 
expansion of the interacting ground state $|\Q_{0}(4)\ket$ [see 
Eq.(\ref{lungo})] by dropping the ones with non-zero direct interaction 
(which are expected to have wrong symmetries to be the ground state). A convenient basis for 
the analytic evaluation of the matrix elements $W_{m,m'}$ and 
$(S[E])_{m,m'}$  (see below) is obtained in terms of the creation operators  
\begin{equation}
f_{1,\sigma}^{\dagger}\equiv \frac{1}{\sqrt{2}}
(c^{\dagger}_{{\bf k}_{3} ,\sigma}+c^{\dagger}_{{\bf k}_{4} ,\sigma}), \quad 
\quad \quad f_{2,\sigma} ^{\dagger} \equiv  
\frac{1}{\sqrt{2}}(c^{\dagger}_{{\bf k}_{3} , \sigma}-c^{\dagger}_{{\bf k}_{4} ,
 \sigma} ).
\end{equation}
The determinantal states 
\begin{equation}
|m_{1} \rangle=f^{\dagger}_{1,\uparrow}\, 
f^{\dagger}_{2,\downarrow}\, | \F_{0}(2)  \rangle , \quad \quad \quad
|m_{2} \rangle=f^{\dagger}_{2,\uparrow}\, 
f^{\dagger}_{1,\downarrow}\, | \F_{0}(2)  \rangle  
\label{emme}
\end{equation}
have projection only on the the first component of  $E_{1}$ and on 
the first component of $T_{2}$, and are mixed 
by the operators $W$ and $S[E]$ in Eq.(\ref{can1}). 
Indeed, the symmetric combination of $|m_{1} \rangle$ and $|m_{2} \rangle$ 
yields $|\q _{1}^{^{1}E_{1}} \rangle$ of Eq.(\ref{pair1}), while the 
antisymmetric combination yields $| \q _{1}^{^{3}T_{2}} \rangle$ 
defined in Eq.(\ref{trip1}). Hence,  the eigenvalue Eq.(\ref{can1}) reduces to: 
\begin{equation}
\left(\begin{array}{cc}
E_{m_{1}}+W_{m_{1}}+F_{m_{1}} & (W_{\rm eff})_{m_{1},m_{2}}  \\
(W_{\rm eff})_{m_{2},m_{1}} & E_{m_{2}}+W_{m_{2}}+F_{m_{2}}
\end{array}\right)\left(\begin{array}{c} a_{m_{1}} \\ 
a_{m_{2}}\end{array}\right)=E\left(\begin{array}{c} a_{m_{1}} \\ 
a_{m_{2}}\end{array}\right),
\label{schmat}
\end{equation}
where we have taken into account that $W^{(d)}=0$ since  
$|m_{1} \rangle$ and $|m_{2} \rangle$ have no-direct interaction and 
that the diagonal part of the effective interaction $W_{\rm eff}$ vanishes 
due to the Pauli principle\footnote{More generally, $(W_{\rm eff})_{m,m}=0$ if 
the $m$ state is a determinantal state. Let 
$m=(\vf_{1,\ua},\vf_{2,\da})$ be a pair index containing two 
spin-orbitals; drawing the diagram for the effective interaction $W_{\rm eff}$ 
with incoming $\vf_{1,\ua}$ and $\vf_{2,\da}$ and outgoing $\vf_{1,\ua}$ and 
$\vf_{2,\da}$ one realizes that one of the two spin-orbitals must be occupied twice.}. 
We performed the sum in Eq.(\ref{fscat}) over the $\a$ states analytically 
using non renormalized $\a$-state energies, which is justified at least in the weak-coupling regime. 
As a consequence of the fact that $(W_{\rm eff})_{m,m}=0$ the forward scattering term 
turns out to be given by  
\begin{equation}
F_{m}=(S[E])_{m,m}=\frac{U^{2}}{8}
\left(\frac{1}{2E}+\frac{1}{E+4t}\right)\equiv f[E]
\label{for11}
\end{equation}
and it is independent of the pair index $m$.
The first-order self-energy $W_{m_{1}}$ is equal to 
$W_{m_{2}}$ and we defer the reader to Section \ref{superc} for the   
proof in a more general case; here we limit to write the final result
\begin{equation}
W_{F}\equiv W_{m_{1}}=W_{m_{2}}=\frac{3U}{32}.
\end{equation}
The off-diagonal matrix elements 
$(W_{\rm eff})_{m_{1},m_{2}}=(W_{\rm eff})_{m_{2},m_{1}}\equiv 
w_{\rm eff}[E]$ are responsible for a spin-flip-like effective 
interaction due to the structure of the states $m_{1}$ and $m_{2}$. 
Performing the sum over the $\a$ states and taking the bare energies 
$E_{\a}$ we get
\begin{equation}
w_{\rm eff}[E]=\frac{U^{2}}{32}\frac{1}{E+4t}.
\end{equation}
Hence, the eigenvectors of Eq.(\ref{schmat}) 
are $\frac{1}{\sqrt{2}}(1,1)$ and  $\frac{1}{\sqrt{2}}(1,-1)$ and 
correspond to the states 
\begin{eqnarray}
|\Psi _{1}^{^{1}E_{1}} \rangle =\frac{1}{\sqrt{2}}
(| m_{1}\rangle +| m_{2}\rangle) \equiv
| \psi _{1}^{^{1}E_{1}} \rangle \otimes |\F_{0}(2) \rangle, \label{sing} \\ 
| \Psi _{1}^{^{3}T_{2}} \rangle =\frac{1}{\sqrt{2}}
(| m_{1}\rangle -| m_{2}\rangle) \equiv
| \psi _{1}^{^{3}T_{2}} \rangle \otimes |\F_{0}(2) \rangle. 
\end{eqnarray}
From the symmetric combination we get the singlet belonging to the irrep $^{1}E_{1}$ 
and the lowest energy $E_{S}$ satisfies the equation
\begin{equation}
-8t+W_{F}+f[E_{S}]+w_{\rm eff}[E_{S}]=E_{S}.
\end{equation}
According to the treatment described in Section \ref{pairmech} we write 
$E_{S}=-8t+W_{F}+f[E_{S}]+\D[E_{S}] $ so that 
\begin{equation}
\Delta[E_{S}] = w_{\rm eff}[E_{S}] 
\end{equation} 
The antisymmetric combination corresponds to the triplet belonging 
to the irrep  $T_{2}$  and the lowest energy $E_{T}$ satisfies the equation 
\begin{equation}
-8t+W_{F}+f[E_{T}]-w_{\rm eff}[E_{T}]=E_{T}.
\end{equation}
In this case $-\D[E_{T}]=w_{\rm eff}[E_{T}]$ and it coincides with the 
modulus of $w_{\rm eff}[E_{S}]$ in standard second-order perturbation 
theory\cite{cibal}, see Fig(\ref{splitting}.$a$). 
\begin{figure}[H]
\begin{center}
	\epsfig{figure=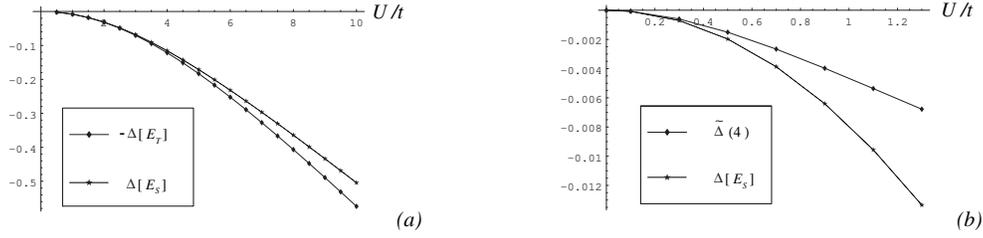,width=13cm}\caption{\footnotesize{
	($a$) $-\D[E_{S}]$ and $\D[E_{T}]$ versus $U/t$. We note that for 
	$U/t>1$ it is evident that $-\D[E_{S}]\neq \D[E_{T}]$. ($b$) 
	Comparison between $\D[E_{S}]$ and $\tilde{\D}(4)$. 
	}}
\label{splitting}
\end{center}
\end{figure}
In Fig.(\ref{splitting}.$b$) it is reported the comparison between 
$\tilde{\D}(4)$, as defined in Eq.(\ref{delta}), and $\D[E_{S}]$ 
obtained from the canonical transformation. 
We have exactly fitted the 8 points for $\tilde{\D}(4)$ and $\D[E_{S}]$  
by using a polynomial of order 7 in $U/t$. The ratio $r$ between the 
quadratic coefficient of $\D[E_{S}]$ and the quadratic coefficient of 
$\tilde{\D}(4)$ is $r=1.00003$, while the linear coefficients are 
essentially zero in both cases. 
We observe that the analytical value $|\Delta[E_{S}]|$ is $\sim  2$ times greater 
than $|\tilde{\D}(4)|$ for $U/t\simeq 1$. This means that the inter-shell interactions 
and the renormalizations of the $\a$-state energies 
have an important weight in determining the right value of 
$\Delta[E_{S}]$. However, what is comfortable is that the 
analytical approach predicts the right trend of the 
binding energy: the singlets feel attraction, while triplet repulsion.
This is the result we need to apply the canonical 
transformation to larger and more physical systems. 

\section{Pairing in nanotube-supercells }
\label{superc}
 
\subsection{Canonical Transformation and Pairing Mechanism for $(N,N)$ 
Nanotubes}

Recently, Potassium\cite{bockrath2} and Lithium\cite{gao} have been 
intercalated in single- and multi-wall carbon nanotubes and a net 
charge transfer was observed between the alkali-metals and the carbon 
atoms. Since the intercalation causes a very small structural 
deformation, we may say that the alkali-metal electrons fill the 
original bands of the nanotube carrying the system away from 
half-filling. In this Section we apply the canonical transformation 
to study the pairing mechanism in the electron-doped ($\ve_{F}>0$) 
armchair nanotubes of length $l=Ld$ and periodic boundary conditions. 
Omitting the band index we choose the 
$m$ states as determinantal eigenstates of the kinetic energy 
$H_{0}$: 
\begin{equation}
c^{\dagger}_{{\bf k}_{1},\uparrow}c^{\dagger}_{{\bf k}_{2},\downarrow}
|\Phi_{0}(n) 
\rangle,\;\;\;\;\varepsilon
({\bf k}_{1}),\varepsilon({\bf k}_{2})>\varepsilon_{F} 
\label{statim}
\end{equation}
where $\e_{F}$ is the
Fermi energy and $|\Phi_{0}(n) \rangle$ is the non interacting Fermi 
{\em sphere} with $n$ particles. The first-order self-energy 
$W_{m}$ is given by 
\begin{equation}
W_{m}=U\sum_{\z=a,b}\left[
\sum_{{\bf k},{\bf k}'}^{\rm occ}|u({\bf k},\z)|^{2}
|u({\bf k}',\z)|^{2}+
\sum_{{\bf k}}^{\rm occ}|u({\bf k},\z)|^{2}
\sum_{i=1}^{2}|u({\bf k}_{i},\z)|^{2}\right]=
U\left[2\left(\frac{n/2}{4NL}\right)^{2}+
\frac{n/2}{4N^{2}L^{2}}\right]\equiv W_{F}
\label{wdiag}
\end{equation}
where $|m\ket=c^{\dag}_{{\bf k}_{1},\ua}c^{\dag}_{{\bf k}_{2},\da}
|\F_{0}(n)\ket$, the sums over ${\bf k}$ and ${\bf k}'$ run over the 
occupied states and we used the fact that $|u({\bf k},\z)|^{2}=1/(4NL)$ 
for any ${\bf k}$ and $\z$. 

We truncate the expansion of the 
interacting ground state $|\Q_{0}(n+2)\ket$ of Eq.(\ref{lungo}) to 
the $\a$ states and a basis for them looks like
\begin{equation}
c^{\dag}_{{\bf k}_{1},\ua}c^{\dag}_{{\bf k}_{2},\da}
c_{{\bf k}_{3},\s}c^{\dag}_{{\bf k}_{4},\s}|\F_{0}(n)\ket,
\;\;\;\;\;\ve({\bf k}_{1}),\ve({\bf k}_{2}),
\ve({\bf k}_{4})>\ve_{F}>\ve({\bf k}_{3}),\;\;
\s=\ua,\da\;.
\label{statia}
\end{equation}

In this scheme of approximation, we  want to obtain the effective interaction 
between the electrons forming the $W=0$ pairs studied in Section \ref{w=0sec}. 
Projecting on the irrep $A_{2}$ the vanishing momentum pair 
$|\q({\bf k}) \rangle\equiv c^{\dag}_{\bf k,\ua}
c^{\dag}_{-{\bf k},\da}|\F_{0}(n)\ket$ we get from Eq.(\ref{fscat}) 
\begin{eqnarray}
\langle \q^{[A_{2}]}({\bf k}) | S[E] | \q^{[A_{2}]}({\bf k}') \rangle =
- 2\d({\bf k}-{\bf k}') \sum_{{\bf p},\n}^{\rm occ} \sum_{{\bf q}}^{\rm emp}  
\frac{\theta(\varepsilon ({\bf k} +{\bf p} -{\bf q}) - \ve_{F} )\;
|U_{\n}({\bf k},{\bf p},{\bf k}+{\bf p}-{\bf q},{\bf q}) |^{\,2}}
{\varepsilon({\bf k}+{\bf p}+{\bf q})-\varepsilon^{\n}({\bf p})+
\varepsilon({\bf q})+\varepsilon({\bf k})-E}+
\nonumber \\ 
 2\sum_{\hat{O} \in C_{2v}} \chi ^{(A_{2})} (\hat{O}) 
\sum_{{\bf p},\n}^{\rm occ} \theta (\varepsilon (\hat{O}{\bf k}' +{\bf k} 
+{\bf p}) - \ve_{F} ) 
\frac{U_{\n}(\hat{O}{\bf k}' +{\bf k} +{\bf p},-{\bf k},\hat{O}{\bf k}',{\bf p}) \, 
U_{\n}({\bf k},{\bf p},\hat{O}{\bf k}'+{\bf k}+{\bf p},-\hat{O}{\bf k}')}
{\varepsilon(\hat{O}{\bf k}'+{\bf k}+{\bf p})-\varepsilon^{\n}({\bf p})+
\varepsilon({\bf k}')+\varepsilon({\bf k})-E}
\label{scattop}
\end{eqnarray}
where we have taken into account Eq.(\ref{statia}) for the $\a$ states 
and we set $E'_{\a}=E_{\a}$, which is justified at least in the 
weak-coupling regime. The sums are over occupied ${\bf p}$ and empty  
${\bf q}$. Since $\ve_{F}>0$, we omitted the band index $+$ of the electron  
eigenenergies $\ve$, while we sum over the band index 
$\n$ of the virtual hole. The vertex  
$U_{\n}$ is given by 
Eq.(\ref{vertex}) where $\n$ refers to the Bloch-function depending 
on ${\bf p}$, while all the other band indices are intended to 
be $+$. The first term on the r.h.s. 
of Eq.(\ref{scattop}) is the forward scattering contribution $F({\bf k})$, 
while the second term  represents the effective interaction 
$W_{\rm eff}({\bf k},{\bf k}')$. 
We may see that standard perturbation theory yields the arithmetic 
mean of the two unperturbed limits $E \to 2\varepsilon({\bf k})$ 
and $E \to 2\varepsilon({\bf k}')$. In particular, the forward scattering 
$F$ in Eq.(\ref{scattop}) coincides with the second-order self-energy of the 
one-particle propagator in the same limit. 

We emphasize that Eq.(\ref{scattop}) is characterized by a symmetry-induced 
quantum mechanical interference of several terms. This interference produces a partial 
cancellation, and the absolute value of the result
is typically much smaller than individual contributions. This means 
that the interaction is dinamically small for $W=0$
pairs: they have no direct interactions (that is $W^{(d)}$, defined in 
Eq.(\ref{selfene}), is zero for $W=0$ pairs), and because of the interference 
the effective interaction is reduced compared to what one could expect  
by  a rough order-of-magnitude estimate. However, the presence of the $theta$
functions and the anisotropy of the integrands prevent a total cancellation.
Substituting Eqs.(\ref{wdiag}-\ref{scattop}) into Eq.(\ref{can1}), we cast the result
in the form of a Cooper-like Schr\"odinger equation
\begin{equation}
\left[2\varepsilon({\bf k}) +W_{F}+F({\bf k})\right]
\,a_{\bf k}+\sum_{{\bf k}'\in {\cal D}/4}\,W_{\rm eff}\,({\bf k},{\bf k}')\,
a_{{\bf k}'} =E a_{\bf k}\;,
\label{cooplike}
\end{equation}
for a self-consistent calculation of $E$ (since $W_{\rm eff}$ and $F$  
are $E$-dependent). The indices ${\bf k}$ and ${\bf k}'$ run 
over $1/4$ of the empty part of the  FBZ and we denoted such a set of 
wavevectors as ${\cal D}/4$. We are interested in the
possibility that $E=2\ve_{F}+W_{F}+F_{\rm min}({\bf k}_{F})+\D$, with a positive binding 
energy $-\D$ of the $W=0$ pair; here $F_{\rm min}({\bf k}_{F})$ is the 
minimum value of $F({\bf k})$ among the ${\bf k}_{F}$-wavevectors on the Fermi 
surface to be determined self-consistently with the eigenvalue $E$.

We have performed numerical estimates of $\Delta$ 
by working on supercells of $2 \, N \times L= N_{C} $ cells. 
Here we solved the Cooper-like equation in a virtually exact way 
for $N$ up to 6 and $L$ up to 32. The results have been reported in Table
V.$a$, V.$b$, V.$c$ and V.$d$ where the hopping parameter $t=1$ eV. 
As a reasonable value, we have taken $U/t=1.7$  which is of the 
correct order of magnitude for  graphite \cite{lopez}\cite{hoffman}. 
In line with our previous finding 
in the small (1,1) cluster, $W=0$ $A_{2}$ singlets show pairing. 
The calculations are performed with the Fermi energy $\ve_{F}$ varying between 
0.5 eV and 1.2 eV (half filling corresponds to $\ve_{F}=0$).  
We see that the binding energy $-\Delta$ of the pairs decreases 
monotonically both with the radius and the length of the tube.  
 
We need to 
work away from half filling in order to operate our mechanism; on the 
other hand, close to half filling the system is a Luttinger liquid
down to extremely low temperatures where a gap could open\cite{sol}.

\begin{center} 
\begin{tabular}{lllll}
\hline
$L$ & $N$  & $-\Delta$ & $-V$ \\
\hline
10 & 2   & 12.0 & 0.43 \\
20 & 2   & 5.7 & 0.38 \\
32 & 2   & 3.6 & 0.35 \\
\hline
\end{tabular} $\quad\quad\quad$
\begin{tabular}{lllll}
\hline
$L$ & $N$ & $-\Delta$ & $-V$ \\
\hline
10 & 4   & 6.3 & 0.37 \\
20 & 4   & 3.1 & 0.38 \\
32 & 4   & 2.0 & 0.35 \\
\hline
\end{tabular} $\quad\quad\quad$
\begin{tabular}{lllll}
\hline
$L$ & $N$  & $-\Delta$ & $-V$ \\
\hline
10 & 6   & 4.0 & 0.39 \\
20 & 6   & 2.4 & 0.47 \\
32 & 6   & 1.5 & 0.38 \\
\hline
\end{tabular} \\
\end{center}
\begin{center}
Table V.$a$. {\footnotesize  Data at $\ve_{F}=0.5$ eV; $-\Delta$ is in meV; 
$V$ is in eV, $t=1$ eV and $U=1.7$ eV.} 
\end{center}
\vspace*{0.1cm}

\begin{center} 
\begin{tabular}{lllll}
\hline
$L$ & $N$  & $-\Delta$ & $-V$ \\
\hline
10 & 2   & 11.6 & 0.42 \\
20 & 2   & 6.0 & 0.40 \\
32 & 2   & 3.7 & 0.36 \\
\hline
\end{tabular} $\quad\quad\quad$
\begin{tabular}{lllll}
\hline
$L$ & $N$ & $-\Delta$ & $-V$ \\
\hline
10 & 4   & 6.0 & 0.40 \\
20 & 4   & 3.0 & 0.40 \\
32 & 4   & 2.0 & 0.43 \\
\hline
\end{tabular} $\quad\quad\quad$
\begin{tabular}{lllll}
\hline
$L$ & $N$  & $-\Delta$ & $-V$ \\
\hline
10 & 6   & 4.2 & 0.38 \\
20 & 6   & 2.2 & 0.36 \\
32 & 6   & 1.1 & 0.30 \\
\hline
\end{tabular} \\
\end{center}
\begin{center}
Table V.$b$. {\footnotesize Data at $\ve_{F}=0.8$ eV; $-\Delta$ is in meV; 
$V$ is in eV, $t=1$ eV and $U=1.7$ eV.} 
\end{center}
\vspace*{0.1cm}

\begin{center} 
\begin{tabular}{lllll}
\hline
$L$ & $N$  & $-\Delta$ & $-V$ \\
\hline
10 & 2   & 11.1 & 0.40 \\
20 & 2   & 6.0 & 0.40 \\
32 & 2   & 3.9 & 0.38 \\
\hline
\end{tabular} $\quad\quad\quad$
\begin{tabular}{lllll}
\hline
$L$ & $N$ & $-\Delta$ & $-V$ \\
\hline
10 & 4   & 6.5 & 0.24 \\
20 & 4   & 2.9 & 0.15 \\
32 & 4   & 1.8 & 0.14 \\
\hline
\end{tabular} $\quad\quad\quad$
\begin{tabular}{lllll}
\hline
$L$ & $N$  & $-\Delta$ & $-V$ \\
\hline
10 & 6   & 4.4 & 0.17 \\
20 & 6   & 3.5 & 0.25 \\
32 & 6   & 1.6 & 0.18 \\
\hline
\end{tabular} \\
\end{center}
\begin{center}
Table V.$c$. {\footnotesize Data at $\ve_{F}=1.0$ eV; $-\Delta$ is in meV; 
$V$ is in eV, $t=1$ eV and $U=1.7$ eV.} 
\end{center}
\vspace*{0.1cm}

\begin{center} 
\begin{tabular}{lllll}
\hline
$L$ & $N$  & $-\Delta$ & $-V$ \\
\hline
10 & 2   & 10.5 & 0.40 \\
20 & 2   & 6.0 & 0.44 \\
32 & 2   & 3.9 & 0.45 \\
\hline
\end{tabular} $\quad\quad\quad$
\begin{tabular}{lllll}
\hline
$L$ & $N$  & $-\Delta$ & $-V$ \\
\hline
10 & 4   & 5.5 & 0.40 \\
20 & 4   & 3.2 & 0.43 \\
32 & 4   & 1.9 & 0.37 \\
\hline
\end{tabular} $\quad\quad\quad$
\begin{tabular}{lllll}
\hline
$L$ & $N$  & $-\Delta$ & $-V$ \\
\hline
10 & 6   & 4.4 & 0.25 \\
20 & 6   & 1.6 & 0.34 \\
32 & 6   & 1.0 & 0.33 \\
\hline
\end{tabular} \\
\end{center}
\begin{center}
Table V.$d$. {\footnotesize Data at $\ve_{F}=1.2$ eV; $-\Delta$ is in meV; 
$V$ is in eV, $t=1$ eV and $U=1.7$ eV.} 
\end{center}
\vspace*{0.1cm}

\subsection{Extrapolation to large $L$}

With supercell sizes $N_{C}>400$ numerical calculations become hard. 
Since we are concerned with the asymptotic behaviour for fixed $N$ and 
$L \to \infty $ and $-\Delta(N,L)$ depends on $N$ and $L$ in a complicated way, 
we need a method to make reliable extrapolations of the numerical results. 
To this end, like in previous work\cite{EPJB1999},\cite{fettes} we define the 
Average Effective Interaction $V$. This is  such that setting in Eq.(\ref{cooplike}) 
$W_{\rm eff}= -\frac{V}{N_{C}}$, with a constant $V>0$ for all ${\bf k}$ and ${\bf k}'$ 
in ${\cal D}/4$, one obtains the correct value of $\D$. In other terms, once the 
binding energy $-\D$ is known by solving Eq.(\ref{cooplike}), the constant $V$ must be chosen 
in such a way that 
\begin{equation}
\frac{1}{V}=\frac{1}{N_{C}}\sum_{{\bf k}\in {\cal D}/4}\,
\frac{1}{[2\varepsilon({\bf k})+F({\bf k})]-[2\ve_{F}+F_{\rm min}({\bf 
k}_{F})]-\D(N,L)}.
\label{uim}
\end{equation}
In Table V.$a$, V.$b$, V.$c$ and V.$d$. we have reported the value 
of $V$ which remains fairly stable around $\approx 0.4$ eV and  does not drop to 
0 in the limit of large $L$. Therefore $V$ must be interpreted as a 
characteristic energy scale of the system which is largely 
independent on the Fermi energy and on the radius. We have 
numerically observed that $|F({\bf k})-F_{\rm min}({\bf k}_{F})|\ll 
-\D$; hence we may extrapolate the asymptotic value of $-\D$ from 
Eq.(\ref{uim}) by dropping the difference $F({\bf k})-F_{\rm min}({\bf 
k}_{F})$ and taking the limit $L\ra \inf$. The domain is extended to 
${\cal D}$ and the result is divided by 4. Since $k_{x}$ is the 
component along the tube axis, we convert the summation into an 
integral: 
\begin{equation}
\frac{1}{V}=\frac{1}{8N}\frac{1}{2\p}
\sum_{k_{y}} \int  dk_{x}  
\frac{ \theta \left(\varepsilon(k_{x},k_{y}) -\ve_{F}  \right)}
{2(\varepsilon(k_{x},k_{y} ) -\ve_{F})-\Delta_{\rm asympt}(N)} \, .
\label{asym}
\end{equation}
with $\Delta_{\rm asympt}(N)=\lim_{L \rightarrow \infty}\D(N,L)$. 
We solved the above equation for the unknown $\Delta_{\rm asympt}(N)$ 
for several values of $N$ and $\ve_{F}$, using a typical value $V\approx 0.4$ eV 
obtained from the calculations in supercells. We found that $\Delta_{\rm asympt}$ is 
strongly dependent on the filling at fixed $N$. Remarkably, there exists an 
{\em optimal doping} at the Fermi energy $\ve_{F}\simeq 1$ eV where 
$\Delta_{\rm asympt}(N)$ is appreciably different from zero, while 
$\Delta_{\rm asympt}$ is strongly suppressed far from it for $N>4$, 
see Fig.(\ref{deltaenef}.$a$). 
\begin{figure}[H]
\begin{center}	
	\epsfig{figure=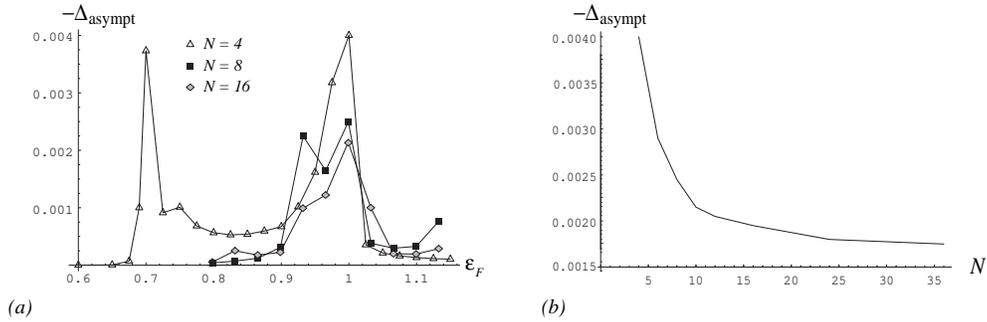,width=13cm}\caption{\footnotesize{Results 
	of the canonical transformation approach with $t=1$ eV and $U=1.7$ eV.
	($a$) $-\Delta_{\rm asympt}$ as a function of the Fermi energy $\ve_{F}$ 
	for $N=4$ (empty triangles), $N=8$ (black boxes) and $N=16$ (grey 
	diamonds). The Fermi energy varies in the range $0.6\div 1.2$. 
	($b$) $-\Delta_{\rm asympt}$ as a function of $N$ for $N$ 
	in the range 4$\div$36 and $\ve_{F}=1$ eV.
	}}
\label{deltaenef}
\end{center}
\end{figure}
The existence of an {\em optimal doping} can be understood by looking 
at the density of states $\rho \, (\varepsilon)$ and at the integrand  
$ 1/ [ \, 2(\varepsilon -\ve_{F})-\Delta_{\rm asympt}\, ] \equiv g(\varepsilon)$  
in Eq.(\ref{asym}). From Fig.(\ref{dens}) we see that $\r(\ve)$ 
has an absolute maximum at $\varepsilon = 1$ eV, which is reminiscent 
of the Van Hove singularity in the graphite sheet, and 
exhibits well-defined oscillations as a function of the 
nanotube radius\cite{ferreira}. 
On the other hand, $g(\varepsilon)$ is peaked at $\ve=\ve_{F}$. Therefore, for 
$\ve_{F} \simeq 1$ eV we have a synergy which leads to an absolute
maximum of $-\Delta_{\rm asympt}$. The prediction of an optimal doping 
follows from Eq.(\ref{asym}) and is therefore largely mechanism-independent. 
A trend similar to the one shown in Fig.(\ref{deltaenef}.b)  was 
reported by Benedict et al.\cite{lorin}, although they considered a 
phonon-driven mechanism.
\begin{figure}[H]
\begin{center}
	\epsfig{figure=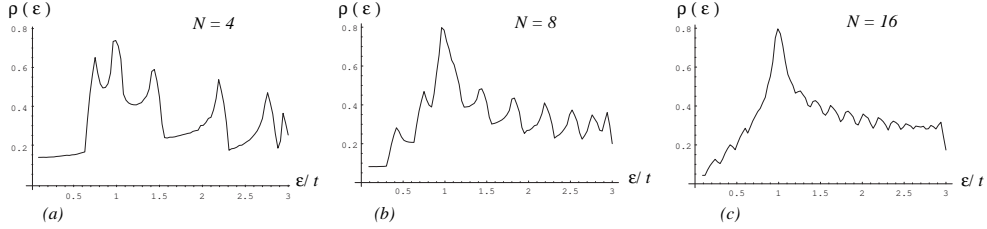,width=13cm}\caption{\footnotesize{
	 Density of states  in the $(4,4)$ ($a$), $(8,8)$ ($b$) and 
	 $(16,16)$ ($c$) nanotube. $\r(\ve)$ is in eV$^{-1}$.
	}}
\label{dens}
\end{center}
\end{figure}

At the {\em optimal doping} we observe that $-\Delta_{\rm asympt}(N)$ 
decreases monotonically as the radius $\sqrt{3}dN/2\p$ of the tube increases, 
see Fig.(\ref{deltaenef}.$b$). However, in the limit of large $N$, 
$\Delta_{\rm asympt}(N)$ remains stable around 1.7 meV and 
may be interpreted as the binding energy of the $W=0$ pair in 
an {\em optimally doped} graphite sheet. By a rough 
order of magnitude estimate, we may say that the superconducting 
critical Temperature predicted by our approach at 
$\ve_{F}=1$ eV (which corresponds to a number of electrons per graphite 
atom of 1.25) is $T_{c}\approx \Delta_{\rm asympt}(\inf)\approx 10\div 20$ K. 
These results may be compared with the available experimental data 
on the alkali-graphite intercalation compounds (GIC). 
There is experimental evidence that the critical Temperature $T_{c}$ 
in alkali-GIC C$_{x}$M (where M is a given alkali metal) grows as $x$ 
decreases\cite{belash}. The number of electrons per graphite atom $f$ 
is related to $x$ by an empirical formula\cite{mizutani} $f=\kappa/x$ 
where $\kappa$ is the fractional charge transfer. A typical value 
for $\kappa$ is 0.5$\div$0.6\cite{jishi} and hence the filling 
$f=1.25$ corresponds to the alkali-GIC's C$_{2}$M. Under 
high-pressure, high metal concentration samples such as 
C$_{6}$K, C$_{3}$K, C$_{4}$Na, C$_{3}$Na, C$_{2}$Na, C$_{2}$Li have 
been synthesized; for C$_{2}$Na the value of $T_{c}$ is 5 K while for 
C$_{2}$Li, $T_{c}$=1.9 K; in both cases the superconducting critical 
Temperature is of the same order of magnitude of 
$\Delta_{\rm asympt}(\inf)$. 
Quite recently Potassium\cite{bockrath2} and Lithium\cite{gao} 
have been intercalated also in single- and multi-wall carbon nanotubes 
up to high concentration (the highest metal concentration was obtained with 
Lithium in C$_{2}$Li). Our mechanism predicts that the binding energy 
of the $W=0$ pairs is bigger in nanotubes than  in graphite 
sheets and this suggests a higher critical Temperature for the 
former, see Fig(\ref{deltaenef}.$b$). This is also supported by the 
measurements of a $T_{c}\approx 15$ K in the 4 Angstrom SWNT 
by Tang {\em et al.}\cite{tang}.

\section{Conclusions}
\label{conc}

Currently, carbon nanotubes superconduct at much lower temperatures than  
high-$T_{c}$ Cuprates and the two kinds of materials are apparently 
quite different. However, symmetry arguments based on  the $W=0$ theorem 
tell us that, despite the obvious differences, part of the story must be 
the same, i.e. by a suitable choice of Dirac's characters the on-site 
Coulomb interaction is utterly turned off.  This produces the singlet 
pairing and constrains the ground state spin-orbital symmetry of the 
interacting system. We presented  analytic expressions  for the 
effective interaction and obtained the binding energy for $(N,N)$ 
armchair nanotubes; in the case $N=1$ we verified these predictions 
numerically and got high-precision agreement. 

Although the results presented in this paper cannot be 
quantitatively compared with the experimetal data, the order of magnitude 
of the pair binding energy agrees with experiment. Furthermore, the 
decreasing of the binding energy with $N$ is suggested by recent 
measurements on nanotubes with diameter of few Angstrom\cite{tang}. 

The paired state we have obtained here is essentially two-dimensional, 
that is the transverse direction is crucial to have a non-Abelian symmetry 
group and hence $W=0$ pairs; the pairing mechanism uses degenerate electronic states that exist 
in 2$d$  away from half filling. This opens up the interesting possibility that two distinct 
superconducting order parameters appear in the phase diagram, if it turns out 
that  close to half-filling there is another one due to a breakdown of 
the Luttinger liquid.

}

\end{document}